# Quantum bit with telecom wave-length emission from a simple defect in Si

Peter Deák,[1,2] Song Li,[2] and Adam Gali[2,3,4,*]

[1] Beijing Computational Science Research Center, Beijing 100193, China

[2] HUN-REN Wigner Research Centre for Physics, P.O. Box 49, H-1525 Budapest, Hungary

[3] Department of Atomic Physics, Institute of Physics, Budapest University of Technology and Economics, Műegyetem rakpart 3., 1111 Budapest, Hungary

[4] MTA-WFK Lendület "Momentum" Semiconductor Nanostructures Research Group

**Abstract**

Spin-to-photon interfaces from defects in silicon hold great promise towards realizing quantum repeaters with the combination of advanced semiconductor and photonics technologies. Recently, controlled creation and erasure of simple carbon interstitial defects have been successfully realised in silicon. This defect has a stable structure near room temperature and emits in the wave-length where the signal loss is minimal in optical fibres used in communication technologies. Our in-depth theoretical characterization confirms the assignment of the observed emission to the neutral charge state of this defect. We find that the emission is due to the recombination of a bound exciton. We also discovered a metastable triplet state that could be applied as a quantum memory. Based on the analysis of the electronic structure of the defect and its similarities to a known optically detected magnetic resonance centre in silicon, we propose that a carbon interstitial can act as a quantum bit and may realize a spin-to-photon interface in CMOS-compatible platforms.

[*]Corresponding author: gali.adam@wigner.hun-ren.hu

**Introduction**

Silicon technology is probably the most advanced area of materials processing, and the readily available and low-cost starting materials ensure the enduring success of silicon-based electronics.

Defects of the silicon crystal, which can be applied as quantum emitters and/or qubits, are receiving increasing attention recently, especially in view of integration with silicon based photonic devices. The defects suggested so far, like the G-center ($C_{Si}$-I-$C_{Si}$),[1,2] the T-centre ($[C_2]_{Si}$+H),[3,4] the W-centre ($I_3$),[5,6,7] and the C-centre ($C_i$+$O_i$),[8,9] are all complexes (where "I" denotes a silicon self-interstitial, the "Si" subscript a substitutional, while the "i" subscript an interstitial impurity), just like the erbium-related centres[10] the structure of which is not even known. Tightly controlled generation of such complexes makes defect engineering very challenging. Therefore, it would be very advantageous to find a simple defect (containing a single impurity) with the appropriate properties. Recently, a defect, with photoluminescence (PL) in the telecom band[11] could be created and erased on demand by applying femtosecond laser pulses with varying the dose of irradiation. Based on the observed frequency, the defect was identified as a single carbon interstitial ($C_i$).[12] The PL of $C_i$ is well known from earlier studies,[5,13,14] and this defect has been amply characterized by deep level transient spectroscopy (DLTS)[15,16,17] and electron spin resonance (ESR)[17,18] as well, providing definitive information about its charge transition levels and geometrical structure. Unfortunately, theoretical calculations could not reproduce these properties so far,[19,20] making the assignment in Ref. [12] somewhat uncertain. It is also an open question whether a single $C_i$ defect can act as a qubit when isolated with the afore-mentioned or other techniques.

Using advanced computational techniques, in this paper we provide a full characterization of the $C_i$ defect, in very good agreement with PL, DLTS, and ESR data, and confirm that the observed emission can really be assigned to its neutral charge state, as a radiative recombination between a bound exciton singlet state and the closed shell localized singlet state. We also find that an optically addressable metastable triplet state also exists, so the defect could be applied as a quantum memory. Comparing it to a known optical detected magnetic resonance centre in silicon (G-centre), we propose that a carbon interstitial could act as a quantum bit and may realize a spin-to-photon interface in CMOS-compatible platforms.

**Results**

The standard implementations of first-principles DFT (density functional theory) contain approximate local- or semi-local exchange functionals (LDA and GGA, respectively), which underestimate the band gap and delocalize defect states. Since defects are usually calculated in big supercells of the host crystal, the application of first-principles many-body methods are as yet computationally prohibitive. Quantitatively correct results may be obtained within (generalized) Kohn-Sham DFT by the application of semi-empirical hybrid exchange functionals, which mix semi-local and non-local (Hartee-Fock-type) exchange. We have shown earlier that the two parameters of the Heyd-Scuseria-Ernzerhof (HSE) exchange functional,[21] i.e., $\alpha$ (for mixing non-local and semi-local exchange) and $\mu$ (to describe electronic screening), can be tuned so that the functional mimics the exact DFT exchange functional,[22,23] i.e., it provides the piece-wise linear behaviour of the total energy as a function of the occupation numbers, with a proper derivative discontinuity at integer values.[24] The latter is equivalent with the reproduction of the exact single-particle band gap. The linearity condition is satisfied, when the generalized Koopmans' theorem (gKT)[25] is fulfilled, i.e., the position of the highest occupied (or lowest unoccupied) Kohn-Sham level matches the ionization energy (or electron affinity), calculated from total energy differences. We have also shown that such optimized HSE($\alpha,\mu$) functionals can yield very accurate results for defects in semiconductors.[26,27,28] Particularly in silicon, the original HSE06 = HSE(0.25,0.20) parametrization[29] is optimal, providing a (0K) band gap of 1.16 eV, in excellent agreement with experiment,[30] and resulting in charge transition levels within 0.1 eV to the measured ones.[26] Therefore, in this study, we apply the HSE(0.25,0.20) functional to calculate the properties of $C_i$, as described in detail in the Methods section.

As it is well known from ESR signals associated with its positive charge state,[17,18] the structure of the carbon interstitial, $C_i$, corresponds to a so-called [001] split interstitial (or dumbbell) configuration, often called $(C-Si)_{Si}^{[001]}$, with the carbon and a silicon atom sharing a lattice site, and giving rise to $C_{2v}$ symmetry. As shown in Fig.1, the threefold coordination results in $sp^2$ hybridization, with pure p-like dangling bonds on both atoms, perpendicular to each other. The C 2p dangling bond is lower in energy than the Si 3p (due to the higher electron affinity of carbon).

Fig.1 also shows the occupations in the neutral ground state and in the possible excited states. Removing or adding an electron preserves the $C_{2v}$ symmetry. Table 1 shows the calculated charge transition levels in comparison to experiment.

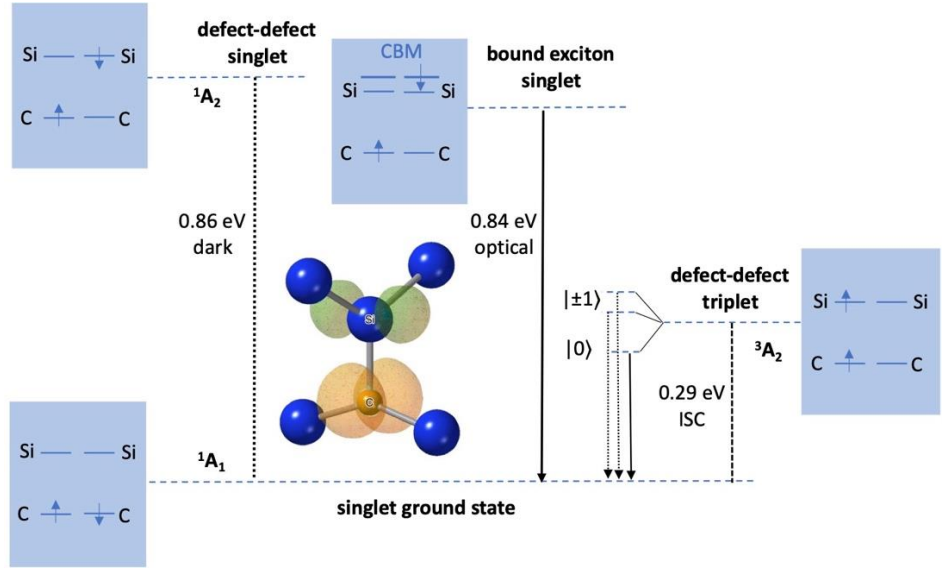

**Fig. 1** Schematic representation of the defect states of $C_i$ (alias $(C-Si)_{Si}^{[001]}$), and their occupation in the ground state and in the dark and bright singlet excited states as well as in the metastable triplet state. The zero-field splitting (ZFS) of the latter is also indicated (with $D$=439 MHz and $E$=38 MHz), from which intersystem crossing (ISC) may occur. The C 2p and Si 3p orbitals are displayed at equal isovalues. (As can be seen, the Si 3p state is a less localized than the C 2p.) The calculated HSE06 values for the ZPL are given, but the arrows showing the transitions are not to scale.

Our calculated formation energy for the neutral charge state (with respect to perfect silicon and diamond) is 3.72 eV, as also found in earlier calculations.[31] We have reported previously[32] the calculated diffusion barrier of $C_i$, via the reorientation mechanism of Ref. [33], to be 0.70 eV (also in good agreement with the observed values between 0.72 and 0.75 eV [17]), which explains the stability of this colour centre below the annealing temperature at ~50 °C.

**Table 1** Calculated and observed charge transition levels. (Note that the gap is 1.12 eV at room temperature (RT), and the acceptor level, measured to be $E_C$ – 0.10 eV at RT, has been converted accordingly with respect to $E_V$. $E_V$ and $E_C$ are the energy of valence and conduction band edges, respectively.)

|  | Present calculation | Experiment (at room temperature) |
| --- | --- | --- |
| E(+/0) | $E_V$ + 0.32 eV | $E_V$ + 0.28 eV; Ref.[17] |
| E(0/-) | $E_V$ + 0.97 eV | $E_V$ + 1.02 eV; Ref.[17] |

We first study the ESR fingerprints of the defect which provide direct information about the localization of the defect wave functions via hyperfine interaction between the electron spin and the nuclear spins. This comparison both verifies the accuracy of our calculations and provides important information about the feasibility of employing this colour centre as a qubit. Table 2 contains the hyperfine tensor in the positive charge state, showing fair agreement with experimental values obtained on a sample enriched with $^{13}$C. This indicates that the applied hybrid functional describes the localization of the defect wave functions well. Hyperfine interaction with $^{29}$Si have not been published to our knowledge but we provide the calculated data in Supplementary Table 1. (We note that another ESR centre was associated with the negatively charged (C-Si)$_{Si}^{[001]}$ defect[17] but no trace of hyperfine interaction with $^{13}$C or $^{29}$Si (within 0.16 T ≈ 4 GHz) was found.) Based on the good agreement of the charge transition levels and the hyperfine data on the $^{13}$C atom in the positive charge state, we conclude that the (C-Si)$_{Si}^{[001]}$ structure is well established, and at the same time the applied method is validated. Next, we focus on the interpretation of the PL spectrum.[5,12]

**Table 2** Calculated and observed hyperfine tensor of the positively charged carbon interstitial, $^{13}$C$_i$.

| $^{13}$C$_i$(+) | A$_{xx}$(MHz) | A$_{yy}$(MHz) | A$_{zz}$(MHz) |
|---|---|---|---|
| Present calculation | 12.4 | 12.0 | 169.44 |
| Experiment; Ref.[18] | 18.71 | 17.60 | 145.7 |

The observed zero-phonon line (ZPL) associated with the C$_i$ defect is at 0.856 eV according to the literature (see, e.g., Ref. [14]), so the corresponding wave length (1448 nm) falls into the telecom region. We note that the defect is susceptible to strain (see e.g., Ref. [18]) and that is likely responsible for the observed variance in the ZPL emissions of the defects in the silicon-on-insulator (SOI) sample[12] which produce a strain field towards the defects in silicon. The observed charge transition levels clearly imply that the negative and positive charge states cannot explain this emission as they would immediately be converted to the neutral state by illumination with higher-than-ZPL energy. We continue the discussion of the emission for the neutral charge state. The ground state is a closed-shell singlet which transforms as the trivial A$_1$. The excited state may be constructed by promoting an electron from the carbon dangling bond to the silicon dangling bond (see Fig. 1). The calculated ZPL energy is 0.86 eV (see Supplementary Note 2 for details) which is close to the experimental value. However, this excited state is dark as transforms as $A_2$ and the optical transition is only allowed by phonon participation which clearly goes against the observed PL spectrum showing a strong ZPL emission. On the other hand, it is intriguing that the donor (+/0) charge transition level, obtained by DLTS is ≈ $E_C$ - 0.87 eV (using the 0K band gap for conversion where $E_C$ is the conduction band minimum), which is only 14 meV higher than the ZPL energy. Thus, the defect may have a bound exciton excited state where the hole is located in the carbon dangling bond orbital and the electron sits on a state split from conduction band minimum (CBM) and the binding energy of the exciton is about 14 meV. Our calculations imply that the dark singlet excited state lies at higher energy than the singlet bound excited state does. Indeed, the calculated ZPL of the bound exciton recombination is 0.84 eV, in good agreement with the measured value of 0.856 eV.

By considering a singlet bound exciton state, the radiative recombination would be allowed in first order. Indeed, the calculated radiative lifetime is 2.83 μs (see Methods). The estimated radiative lifetime is relatively long because the excited state is delocalized whereas the ground state is localized, which results in a relatively weak optical transition dipole moment (0.96 D). Nevertheless, the spectrum does show a significant ZPL emission according to our simulations (see Fig. 2). The phonon sideband in the PL spectrum can be well explained by the ion relaxations going from the equilibrium geometry of the positively charged defect to that of the neutral defect. The features can be identified as the phonon modes of the Si crystal as the positively charged defect produces a larger tensile strain than the neutral one does. The calculated Huang-Rhys (HR) factor is 2.88. Based on these results, we establish the origin of the PL signal as an optical transition from a bound exciton state to the closed-shell singlet ground state of the neutral $C_i$ defect.

We now discuss the presence of optically inactive or dark states and their roles. We assume that the dark singlet level may fall above the ionisation energy of the neutral defect. In the opposite scenario, the dark state would significantly reduce the brightness of the defect where the bright state may be activated by elevating the temperature to occupy the bright singlet state. However, it was observed in a recent study[12] that the emission intensity is decreasing with raising the temperature which clearly indicates that the dark state's level should lie above the bright state's level. The temperature dependence of the PL intensity can be rather explained by the thermal ionisation of the bound exciton excited state.

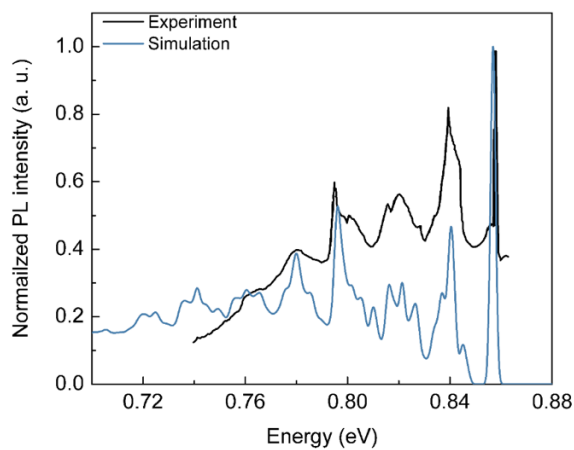

**Fig. 2** The calculated and the observed[13] PL spectra, including the phonon sidebands. Here we use the neutral and the positive charge state's geometries as ground and excited states, respectively, to generate the phonon sideband of the PL spectrum. The simulated spectrum is aligned to the experimental one at the zero-phonon line for direct comparison of the experimental and simulated phonon sidebands. We note that the experimental spectrum sits on a tail of a broad background that is not present in the simulated spectrum.

We have also explored a triplet manifold which is a metastable state of the defect. This may be generated by promoting an electron from the carbon dangling bond orbital to the silicon dangling bond orbital with spin flip. The corresponding $^3A_2$ level lies 0.29 eV above the ground state. The triplet state keeps the $C_{2v}$ symmetry. The spin levels split by the orthorhombic crystal field with $D$=439.3 MHz and $E$=37.9 MHz. The $^{13}$C and $^{29}$Si hyperfine couplings are characteristic to the defect as the spin density is mostly localized on both dangling bonds (see Supplementary Table 3).

**Discussion**

Silicon is a very promising platform for hosting near-infrared single photon emitters and quantum bits realized by fluorescent point defects.[34] Stable defects with a single impurity may be preferred over complexes where the latter often require difficult defect engineering protocols to create them even with relatively low yield. The recently demonstrated programmable creation of carbon interstitial defects[12] gives hope that telecom wave length single photon emitters can be generated on demand in a given density with good scalability.

The calculated radiative lifetime is 2.83 $\mu$s whereas the observed PL lifetime upon 532-nm pulsed laser excitation is 3-8 ns.[12] The PL lifetime inherits both radiative and non-radiative components where various non-radiative processes may occur such as internal conversion, intersystem crossing or ionisation. Since the excitation energy of a 532-nm laser (2.33 eV) is much larger than the band gap of Si and stable positive and negative charge states of the $C_i$ defect exist, it is likely that ionisation and re-ionisation processes occur depending on the excitation power. Indeed, the PL intensity depends non-linearly on the laser power[12] which is a clear signature that photo-ionisation processes take place in the $C_i$ defect. Our results imply that the quantum yield of excitation is about 0.1% at the given experimental conditions in Ref. [12]. As a consequence, observation of single defect emission requires a photonics structure at the ZPL wavelength in order to significantly enhance the PL intensity. This concept was already demonstrated for the G-centre[35] and the T-centre,[36] so it is viable towards $C_i$ emitters too.

In order to turn the single photon emitter into a quantum repeater unit, a quantum memory should be associated with the defect. We found a metastable triplet state which can act as a quantum memory. Decay from the metastable triplet state to the singlet ground state occurs via intersystem crossing (ISC). According to group theory, the $|^3A_2, m_S=0\rangle$ state is linked to the ground state $^1A_1$, whereas the ISC is forbidden from $|^3A_2, m_S=\pm 1\rangle$ in the first order. This means that the lifetime of $|^3A_2, m_S=0\rangle$ is shorter than that of $|^3A_2, m_S=\pm 1\rangle$. This electronic structure and selection rules of ISC strongly resembles those of the G-centre in Si (Ref. [2]) for which optically detected magnetic resonance (ODMR) of the electron spin was already demonstrated.[37] We provide the hyperfine coupling parameters in Supplementary Note 3 that should be assessed in future experiments for identification of the centre. By applying a microwave pulse resonant with the $|^3A_2, m_S=0\rangle$ and $|^3A_2, m_S=+1\rangle$ or $|^3A_2, m_S=-1\rangle$ levels (see Fig. 1), a contrast in the PL emission of the $C_i$ defect is expected upon resonant microwave transition. Proximate nuclear spins may be applied as quantum registers to store the quantum information which was encoded in the electron spin. On the other hand, too strong hyperfine interaction between the electron spin and nuclear spins has a detrimental effect on the coherence of the electron spin. According to our calculations, the hyperfine constants of $^{29}$Si with $I=1/2$ nuclear spin in the second and third neighbour shells are all in the order of 10 MHz which could significantly shorten the coherence time of the electron spin. Therefore, it is critical that the defect should be created in isotope engineered silicon with significant reduction of $^{29}$Si content (to about 0.5% from the natural abundant 4.5%) so that the probability of finding one $^{29}$Si around the defect up to the third neighbour shell could be reduced below 1%. The farther situated $^{29}$Si nuclear spins can be safely applied as quantum memories where the quantum information from the electron spin to the nuclear spin can be realised by the Landau-Zener effect or other techniques.

In conclusion, our findings on the electronic structure and magneto-optical properties of a simple carbon interstitial defect in silicon establish a spin-to-photon interface with ZPL emission at the wave length compatible with the fibre optics based communication in the most mature optoelectronics platform. Our analysis revealed that it is critical to engineer this colour centre into a photonics structure for observation of single defect emission, and the ODMR experiments and quantum memory operation can be realised in $^{28}$Si isotope enriched silicon host.

**Methods**

HSE(0.25,0.20) calculations were carried out with the Vienna *ab-initio* simulation package VASP 5.4, using the projector augmented wave method,[38,39,40] and a plane wave cutoff of 420 (840) eV for the wave function (charge density). Defects were modelled in a 512-atom supercell (4×4×4 multiple of the conventional Bravais-cell), and all atoms were allowed to relax in a constant volume till the forces were below 0.01 eV/Å. The Γ-point approximation was used for Brillouin-zone sampling. The lattice constant was taken from our earlier HSE06 work[26] to be 5.4307 Å (in good agreement with experiment). Total energies of charged systems have been calculated by applying the SCPC method for charge correction.[41] The zero phonon line (ZPL) was obtained as the energy difference of the relaxed ground and excited states, the latter calculated with constrained occupation or $\Delta$SCF method. To obtain the correct ZPL for the singlet-to-singlet transition, an exchange correction was applied.[42] In case of the bound exciton, a band-filling correction was also used.[43] The spectrum of the phonon replicas were computed by the generating function method,[44] based on vibration calculations using the Perdew-Burke-Ernzerhof (PBE)[45] functional. The radiative lifetime was calculated with the inverse of $\Gamma_{\text{rad}} = \frac{n_D E_{\text{ZPL}}^3 \mu^2}{3\pi \varepsilon_0 c^3 \hbar^4}$ (see Ref. [46]), where the optical transition dipole moment $\mu$ was calculated by taking the respective Kohn-Sham wave functions representing the electronic ground and excited states, and $n_D$=3.485 is the refractive index of silicon at 1.45 μm or ≈0.85 eV (see Ref. [47]), $E_{\text{ZPL}}$=0.856 eV is the observed energy of the ZPL, $\varepsilon_0$ is the dielectric permittivity of vacuum, $c$ is the speed of light and $\hbar$ is the reduced Planck-constant.

The hyperfine tensors and zero-field splitting tensor were calculated as implemented in VASP[48,49] where the zero-field splitting tensor algorithm was implemented by Martijn Marsman.

**Data Availability**

The authors declare that the main data supporting the findings of this study are available within the paper and its Supplementary files. Part of the source data is provided in this paper. The data that support the findings of this study are available from the corresponding author upon reasonable request.

**Code availability**

The codes that were used in this study are available upon request to the corresponding author.

**Acknowledgements**

A.G. acknowledges the EU HE projects QuMicro (Grant No. 101046911) and SPINUS (Grant No. 101135699). This research was supported by the Ministry of Culture and Innovation and the National Research, Development and Innovation Office within the Quantum Information National Laboratory of Hungary (Grant No. 2022-2.1.1-NL-2022-00004). We acknowledge KIFÜ for awarding us access to high-performance computation resource based in Hungary. Open access funding provided by HUN-REN Wigner Research Centre for Physics.

**Contributions**

S.L. and P.D. carried out the DFT calculations under the supervision of A.G. P.D. and A.G. analysed the results. All authors contributed to the discussion and writing of the paper. A.G. conceived and led the entire scientific project.

**Corresponding author**

Correspondence to [Adam Gali](Adam Gali).

**Competing interests**

The authors declare no competing interests.

**References**

[1] Song, L. W. Zhan, X. D., Benson, B. W. and Watkins, G. D. Bistable interstitial-carbon —substitutional-carbon pair in silicon. *Phys. Rev. B* **42**, 5765 (1990).
[2] Udvarhelyi, P. Somogyi, B. Thiering, G. and Gali, A. Identification of a Telecom Wavelength Single Photon Emitter in Silicon. *Phys. Rev. Lett.* **127**, 196402 (2021).
[3] Safonov, A. N., Lightowlers, E. C., Davies, G., Leary, P., Jones, R., and Öberg, S. Interstitial-Carbon Hydrogen Interaction in Silicon. *Phys. Rev. B* **77**, 4812 (1996).
[4] Dhaliah, D., Xiong, Y., Sipahigil, A., Griffein, S. M., and Hautier, G. First-principles study of the T center in silicon. *Phys. Rev. Mater.* **6**, L053201 (2022).
[5] Davies, G. The optical properties of luminescence centers in silicon. *Phys. Reports* **176**, 83 (1989).

[6] Tan, J. Davies, G. Hayama, S. Harding, R. and Wong-Leung, J. Ion implantation effects in silicon with high carbon content characterized by photoluminescence, *Physica B* **340–342** 714 (2003).

[7] Baron, Y. Durand, A. Herzig, T. Khoury, M. Pezzagna, S. Meijer, J. Robert-Philip, I. Abbarchi, M. Hartmann, J.-M. Reboh, S. Gérard, J.-M. Jacques, V. Cassabois, G. Dréau, A. Single G centres in silicon fabricated by co-implantation with carbon and proton. *Appl. Phys. Lett.* **121**, 084003 (2022).

[8] Davies, G. Carbon-related processes in crystalline silicon. *Mater. Sci. Forum,* **38**, 151 (1989).

[9] Udvarhelyi, P., Pershin, A., Deák, P., and Gali, A. An L-band emitter with quantum memory in silicon. *npj Comput. Mater.* **8**, 262 (2022).

[10] Kenyon, A. J. Erbium in silicon. *Semicond. Sci. Technol.* **20**, R65 (2005).

[11] https://www.thefoa.org/tech/ref/basic/SMbands.html

[12] Jhuria, K., Ivanov,V., Polley, D., Liu, W., Persaud, A., Zhiyenbayev, Y., Redjem, W., Qarony, W., Parajuli, P., Qing, J., Gonsalves, A. J., Bokor, J., Tan, L. Z., Kanté, B. and Schenkel, T. Programmable quantum emitter formation in silicon. *arXiv:2307.05759* (2023).

[13] Thonke, G., Teschner, A., and Sauer, R. New photoluminescence defect spectra in silicon irradiated at 100 K: Observation of interstitial carbon? *Sol. State Commun.* **61**, 241 (1987).

[14] Woolley, R., Lightowlers, E. C., Tipping, A. K., Clayburn, M., and Newman, R. C. Electronic and vibrational absorption of interstitial carbon in silison. *Mater. Sci. Forum* **10-12**, 929 (1986).

[15] Lee, Y. H., Cheng, L. J., Gerson, J. D., Mooney, P. M., and Corbett, J. W. Carbon interstitial in electron-irradiated silicon. *Sol. State Commun.* **21**, 109 (1977).

[16] Kimmerling, L. C., Blood, P. and Gibson, W. M. *IOP Conf. Proc. Ser.* **46**, 273 (1979).

[17] Song, L. W. and Watkins, G. D. EPR identification of the single-acceptor state of interstitial carbon in silicon. *Phys. Rev. B* **42**, 5759 (1990).

[18] Watkins, G. D. and Brower, K. L., *Phys. Rev. lett.* 36, 1329 (1976).

[19] Leary, P., Jones, R., Öberg, S., and Torres, V. J. B. Dynamic properties of interstitial carbon and carbon-carbon pair defects in silicon. *Phys. Rev. B* **55**, 2188 (1997).

[20] Wang, H., Chroneos, A., Londos, C. A., Sgourou, E. N., and Schwingenschlögl, U. Carbon related defects in irradiated silicon revisited. *Sci. Reports* **4**, 4909 (2014).

[21] Heyd, J., Scuseria, G. E., and. Ernzerhof, M. Hybrid Functionals Based on a Screened Coulomb Potential. *J. Chem. Phys.* **118**, 8207 (2003).

[22] Deák, P., Ho, Q. D., Seemann, F., Aradi, B., Lorke, M. and Frauenheim, T. Choosing the correct hybrid for defect calculations: a case study on intrinsic carrier trapping in ß-$Ga_2O_3$. *Phys. Rev. B* **95**, 075208 (2017).

[23] Deák, P., Lorke, M., Aradi, B., and Frauenheim, T. Optimized hybrid functionals for defect calculations in semiconductors. *J. Appl. Phys.* **126**, 130901 (2019).

[24] Kaplan, A. D., Levy, M. and Perdew, J. P. The Predictive Power of Exact Constraints and Appropriate Norms in Density Functional Theory, *Ann. Rev. Phys. Chem.* **74**, 193 (2023).

[25] Lany, S. and Zunger, A. Polaronic hole localization and multiple hole binding of acceptors in oxide wide-gap semiconductors, *Phys. Rev. B* **80**, 085202 (2009).

[26] Deák, P., Aradi, B., Frauenheim, T., Janzén, E. and Gali, A. Accurate defect levels obtained from the HSE06 range-separated hybrid functional. *Phys. Rev. B* **81**, 153203 (2010).

[27] Han, M., Zeng, Z., Frauenheim, T. and Deák, P. Defect physics in intermediate-band materials: Insights from an optimized hybrid functional. *Phys. Rev. B* **96**, 165204 (2017).

[28] Han, M., Deák, P., Zeng, Z. and Frauenheim, T. Possibility of Doping $CuGaSe_2$ n-Type by Hydrogen. *Phys. Rev. Appl.* 15, 044021 (2021).

[29] Krukau, A. V., Vydrov, O. A., Izmaylov, A. F. and Scuseria, G. E. Influence of the Exchange Screening Parameter on the Performance of Screened Hybrid Functionals. *J. Chem. Phys.* **125**, 224106 (2006).

[30] Cardona, M. and Thewalt, M. L. Isotope effects on the optical spectra of semiconductors. *Rev. Mod. Phys.* **77**, 1174 (2005).

[31] Zirkelbach, F., Stritzker, B., Nordlund, K., Lindner, J. K. M., Schmidt, W. G., and Rauls, E. Combined *ab initio* and classical potential simulation study on silicon carbide precipitation in silicon. *Phys. Rev. B* **84**, 064126 (2011).

[32] Deák, P., Udvarhelyi, P., Thiering, G., and Gali, A. The kinetics of carbon pair formation in silicon prohibits reaching thermal equilibrium. *Nat. Commun.* **14**, 361 (2023).

[33] Capaz, R. B. Dal Pino Jr., A. and Joannopoulos, J. D. Identification of the migration path of interstitial carbon in silicon. *Phys. Rev. B* **50**, 7439 (1994).

[34] Zhang, G., Cheng, Y., Chou, J.-P., and Gali, A. Material platforms for defect qubits and single-photon emitters. *Appl. Phys. Rew.* **7**, 031308 (2020).

[35] Prabhu, M., Errando-Herranz, C., De Santis, L., Christen, I., Cgen, Ch., Gerlach, C., and Englund, D. Individually addressable and spectrally programmable artificial atoms in silicon photonics. *Nature Commun.* **14**, 2380 (2023).

[36] Higginbottom, D. B., et al. Optical observation of single spins in silicon. *Nature* **607**, 2566 (2022).

[37] O'Donnell, K. P., Lee, K. M., and Watkins, G. D. Origin of the 0.97 eV luminescence in irradiated silicon. *Physica B* **116**, 258 (1983).

[38] Kresse, G. and Hafner, J. Ab initio molecular-dynamics simulation of the liquid-metal–amorphous-semiconductor transition in germanium. *Phys. Rev. B* 49, 14251 (1994).

[39] Kresse, G. and Furthmüller, J. Efficient iterative schemes for ab initio total-energy calculations using a plane-wave basis set. *Phys. Rev. B* **54**, 11169 (1996).

[40] Kresse, G. and Joubert, D. From ultrasoft pseudopotentials to the projector augmented-wave method. *Phys. Rev. B* **59**, 1758 (1999).

[41] Chagas de Silva, M. Lorke, M. Aradi, B. Farzalipour Tabriz, M. Frauenheim, T. Rubio, A, Rocca, D. and Deák, P. Self-consistent potential correction for charged periodic systems, *Phys. Rev. Lett.* **126**, 076401 (2021).

[42] Ziegler, T., Rauk, A., and Baerends, E. J. On the calculation of multiplet energies by the hartree-fock-slater method. *Theoret. Chim. Acta* **43**, 261 (1977).

[43] Lany, S. and Zunger, A. Assessment of correction methods for the band-gap problem and for finite-size effects in supercell defect calculations: Case studies for ZnO and GaAs. *Phys. Rev. B* **78**, 235104 (2008).

[44] Alkauskas, A., Buckley, B. B., Awschalom, D. D., & Van de Walle, C. G. First-principles theory of the luminescence lineshape for the triplet transition in diamond NV centres. *New J. Phys.*, 16, 073026 (2014).

[45] Perdew, J. P. Burke, K. Ernzerhof, M. Generalized Gradient Approximation Made Simple. *Phys. Rev. Lett.* **77**, 3865 (1996).

[46] Li, S., Thiering, G., Udvarhelyi, P., Ivády, V., and Gali, A. Carbon defect qubit in two-dimensional $WS_2$. *Nature Commun.* **13**, 1210 (2022).

[47] Green, M. A. Self-consistent optical parameters of intrinsic silicon at 300 K including temperature coefficients. *Solar Energy Mater, & Solar Cells* **92**, 1305 (2008).

[48] Bodrog, Z. and Gali, A. The spin–spin zero-field splitting tensor in the projector-augmented-wave method. *J. Phys. Condens. Matter.* **26**, 015305 (2013).

[49] Szász, K., Hornos, T., Marsman, M., and Gali, A. Hyperfine coupling of point defects in semiconductors by hybrid density functional calculations: The role of core spin polarization. *Phys. Rev. B* **88**, 075202 (2013).